\documentclass[
aps,prb,twocolumn,amsmath,amssymb,floatfix,
superscriptaddress,longbibliography,footinbib
]{revtex4-2}

\usepackage{bm}
\usepackage{float}
\usepackage{enumerate}

\usepackage{graphicx}
\usepackage{epstopdf}

\usepackage{dcolumn}
\usepackage{latexsym}
\usepackage{verbatim}
\usepackage{comment}
\usepackage{physics}
\usepackage{stmaryrd}
\usepackage{color}
\usepackage{epstopdf}
\usepackage{grffile}
\usepackage{lipsum}
\usepackage{enumitem}
\usepackage[toc,page]{appendix}
\usepackage[dvipsnames]{xcolor}
\usepackage{soul}

\DeclareGraphicsExtensions{.eps}

\newcommand{\beq}{\begin{equation}}
\newcommand{\eeq}{\end{equation}}
\newcommand{\bea}{\begin{eqnarray}}
\newcommand{\eea}{\end{eqnarray}}
\newcommand{\ben}{\begin{eqnarray*}}
\newcommand{\een}{\end{eqnarray*}}

\newcommand{\bfig}{\begin{figure}}
\newcommand{\efig}{\end{figure}}

\usepackage[
colorlinks=true,
urlcolor=blue,
citecolor=blue,
linkcolor=blue
]{hyperref}

\begin{document}
\title{Multiple windows of eigenstate-selective delocalization in a non-Hermitian quasiperiodic lattice}
\author{Souvik Roy}
\email{souvikroy138@gmail.com}
\affiliation{School of Basic Sciences, Indian Institute of Technology Bhubaneswar, Odisha 752050, India}

\author{Santanu K. Maiti}

\email{santanu.maiti@isical.ac.in}
	
\affiliation{Physics and Applied Mathematics Unit, Indian Statistical Institute, 203 Barrackpore Trunk Road, Kolkata-700 108, India}

\date{\today}

\begin{abstract}
We investigate localization in a non-Hermitian quasiperiodic lattice with a real quasiperiodic onsite potential of strength $W$ and a staggered imaginary quasiperiodic modulation of strength $\lambda$. Using the inverse
participation ratio (IPR), normalized participation ratio (NPR), fractal dimension $D_2$, transfer-matrix Lyapunov exponent, and wave-packet dynamics, we find a strongly non-monotonic and state-dependent response. At fixed $\lambda$, varying $W$ yields two distinct delocalization windows,
whereas varying $\lambda$ at fixed $W$ produces a single re-entrant delocalization window. These regimes arise within selected eigenstate sectors, where localized states coexist with extended or critical states, leading to mobility-edge-like behavior. The corresponding suppression of the Lyapunov exponent and enhancement of the dynamical RMS spreading are
consistent with the IPR, NPR, and $D_2$ results. Finite-size scaling confirms the persistence of the re-entrant features, while the complex-energy spectrum resolves their state-selective structure. The resulting phase boundaries delineate distinct localization regimes in the
$(W,\lambda)$ plane. Our results show that the interplay between quasiperiodicity and a staggered non-Hermitian modulation sharing the same quasiperiodic spatial structure can generate re-entrant, state-dependent localization, with implications for localization and dynamics in non-Hermitian quasiperiodic lattices.
\end{abstract}

\maketitle

\section{Introduction}

Localization of quantum states in spatially nonuniform media is a fundamental problem in condensed matter physics, originating from Anderson's seminal demonstration that disorder can inhibit quantum transport through interference among multiple scattering processes~\cite{ander1,ander2,ander3,ander4}. The concept has subsequently been explored across diverse settings, including electronic, optical, acoustic, and ultracold-atom systems. Although randomness is the conventional source of localization, spatially deterministic modulations can produce equally nontrivial localization behavior. Quasiperiodic lattices~\cite{dis1, dis6, dis13} are particularly useful in this context, as they retain long-range spatial structure while lacking the translational periodicity of a crystal, placing them between periodic order and random disorder. The Aubry-Andr\'e-Harper (AAH) model~\cite{dis3,dis4,dis5, dis7, dis8,dis9,dis10,dis11,dis14} provides a prototypical realization, with its self-duality yielding a transition between extended and localized eigenstates~\cite{dis15,dis16,dis17,dis18,dis19,dis20,dis21,dis23}. Extensions of the AAH framework have further revealed mobility edges~\cite{me1,me2,sanchu}, multifractal eigenstates, and other unconventional localization regimes, demonstrating that the localization properties can depend sensitively on the spatial form of the modulation, lattice structure, and hopping amplitudes.

Non-Hermiticity~\cite{nh1,nh2,nh3,nh4,nh5,nh6,nh7,nh8} has emerged as another important direction in quantum physics, particularly over the past decade. Such Hamiltonians provide an effective framework for describing systems coupled to their environment, while also enabling phenomena that do not arise in conventional Hermitian settings. This has led to the discovery and development of a broad range of effects, including exceptional points, parity-time symmetry breaking, non-Bloch band descriptions, the non-Hermitian skin effect~\cite{nh30,nh32,nh34,nh31,nh35,nh12}, unconventional spectral topology~\cite{nh36,nh37,nh54,nh57,nh56,nh33,nh41,nh28,nh16}, and anomalous transport~\cite{svkr1,svkr2}. Advances in experimental~\cite{expnh1} platforms such as photonic and electrical lattices, mechanical metamaterials, and ultracold atoms have made these effects increasingly accessible and have driven substantial theoretical and experimental activity. An important issue is how non-Hermiticity modifies the localization of quantum states. Unlike a conventional Hermitian perturbation, a complex onsite potential or nonreciprocal coupling can alter both the spatial structure of eigenstates and their spectral properties. Studies~\cite{spadhi1,spadhi2,spadhi3,hsu} of non-Hermitian lattices have therefore revealed substantial modifications of localization transitions, including changes in mobility-edge behavior, nonreciprocal propagation, and critical regimes without direct Hermitian analogues. These developments raise the broader question of how quasiperiodic localization is reshaped when deterministic spatial modulation is combined with a structured non-Hermitian potential.

The combination of quasiperiodic order with non-Hermiticity has consequently become an important setting for exploring localization beyond the conventional Hermitian picture. Non-Hermitian quasiperiodic models have demonstrated that complex onsite modulations and nonreciprocal couplings can reshape localization boundaries, alter the structure of critical states, and produce mobility-edge-like and re-entrant behavior~\cite{nh11,nh22,nh10,nh9, mainre}. These effects indicate that localization is governed not only by the strength of the quasiperiodic modulation but also by its spatial relation to the non-Hermitian component. A particularly relevant question is therefore how a spatially structured non-Hermitian potential modifies localization when combined with quasiperiodic order and dimerized hopping. In this setting, the possibility of state-selective transitions and intermediate regimes containing different localization characters remains insufficiently explored. Addressing these issues is essential for understanding how lattice structure and spatial non-Hermiticity jointly determine the localization properties of quasiperiodic systems.

Motivated by these considerations, we investigate a non-Hermitian quasiperiodic lattice with a real quasiperiodic onsite potential and a staggered imaginary quasiperiodic modulation. The combined presence of quasiperiodicity, and spatially structured non-Hermiticity introduces competing mechanisms that can substantially modify the localization properties. We characterize the resulting behavior using complementary measures, including the IPR, NPR, fractal dimension $D_2$, Lyapunov exponent, and eigenstate-resolved localization diagnostics. These measures allow us to distinguish changes involving the spectrum as a whole from those restricted to particular eigenstate sectors, while the complex-energy spectrum and wave-packet dynamics provide independent perspectives on the same localization phenomena.

Our central finding is a strongly non-monotonic and asymmetric localization response to the two control parameters. At fixed non-Hermitian strength, increasing the quasiperiodic potential can drive the system through two separate delocalized regimes, with localization recovered between them, whereas varying the non-Hermitian strength at fixed quasiperiodic potential produces a single re-entrant delocalization regime. Importantly, these successive changes are not uniform across the spectrum but are concentrated within selected eigenstate sectors, leading to an intermediate regime with mobility-edge-like characteristics. This state-selective re-entrant behavior demonstrates that the spatial structure of the non-Hermitian modulation, together with quasiperiodicity, can qualitatively reshape the localization phase structure.

The remainder of the paper is organized as follows. In Sec.~II, we introduce the model Hamiltonian and define the localization measures used throughout the analysis. Section~III presents the main numerical results, including the averaged and eigenstate-resolved localization properties, spectral characteristics, critical behavior, and phase diagram. Section~IV summarizes the principal findings and discusses their implications. Additional analysis supporting the choice and construction of the model is provided in Appendices~A and~B.

\section{Model and Theoretical Formulation}
In this section, we introduce the tight-binding Hamiltonian of the proposed one-dimensional non-Hermitian quasiperiodic system and establish the theoretical framework used to characterize its localization properties. We specify the hopping and onsite terms of the model and provide the definitions and computational prescriptions for the localization diagnostics employed throughout the work.

We study the localization properties of a one-dimensional quasiperiodic lattice in the presence of spatially structured non-Hermiticity. The system combines nearest-neighbor hopping with a real quasiperiodic onsite potential and a staggered imaginary quasiperiodic modulation. We impose periodic boundary conditions and consider a lattice of $L$ sites. The corresponding single-particle Hamiltonian is written as
\begin{equation}
H=
\sum_{n=1}^{L}
\left[
t\left(c_n^\dagger c_{n+1}
+c_{n+1}^\dagger c_n\right)
\right]
+
\sum_{n=1}^{L}V_n c_n^\dagger c_n,
\; c_{L+1}\equiv c_1,
\end{equation}
where the nearest-neighbor hopping (NNH) amplitude $t$ is fixed throughout, while the onsite potential takes the form
\begin{equation}
V_n=
W\cos(2\pi b n)
+i\lambda(-1)^n
\sin(2\pi b n),
\end{equation}
with $b=(1+\sqrt{5})/2$. Here, $W$ controls the real quasiperiodic contribution, while $\lambda$ sets the amplitude of the staggered imaginary term. For each choice of these parameters, we obtain the complex eigenenergies and right eigenstates by diagonalizing the single-particle Hamiltonian,
\begin{equation}
H|\psi_n\rangle=E_n|\psi_n\rangle ,
\qquad
E_n=E_n^{\mathrm R}+iE_n^{\mathrm I}.
\end{equation}
For the energy-resolved analysis, the eigenstates are indexed according to increasing $\operatorname{Re}(E_n)$.

\subsection{Localization diagnostics}

To quantify the spatial character of the eigenstates, we employ several complementary diagnostics based on the normalized right eigenvectors. Let $\ket{\psi_n}$ denote the $n$th right eigenstate, with site amplitudes $\psi_j^{(n)}$ satisfying
\begin{equation}
\sum_{j=1}^{L}|\psi_j^{(n)}|^2=1.
\end{equation}
The degree of spatial localization is first characterized by the inverse participation ratio (IPR)~\cite{IPR},
\begin{equation}
\mathrm{IPR}_n=
\sum_{j=1}^{L}|\psi_j^{(n)}|^4.
\end{equation}
This quantity measures the spatial concentration of an eigenstate through the fourth moment of its probability amplitudes. For a state extended over the entire lattice, the weight is distributed over many sites and $\mathrm{IPR}_n$ scales as $L^{-1}$, approaching zero in the thermodynamic limit. In contrast, a state localized on a finite number of sites retains a finite IPR as $L$ increases, with the limiting value reaching $\mathrm{IPR}_n=1$ for a state completely concentrated on a single site. Thus, $\mathrm{IPR}_n\rightarrow0$ characterizes an extended state, whereas $\mathrm{IPR}_n\rightarrow1$ represents the maximally localized limit, making the IPR a direct measure of the spatial concentration of an individual eigenstate.

For convenience, we also use the normalized participation ratio (NPR)~\cite{NPR},
\begin{equation}
\mathrm{NPR}_n=
\frac{1}{L\,\mathrm{IPR}_n}.
\end{equation}
The NPR provides a complementary measure of the spatial extent of an eigenstate. For an ideally extended state with approximately uniform weight over all $L$ sites, $\mathrm{IPR}_n\sim L^{-1}$ and $\mathrm{NPR}_n$ approaches unity. In contrast, for a localized state whose weight remains confined to a finite spatial region, the IPR remains finite and $\mathrm{NPR}_n\sim L^{-1}$ vanishes in the thermodynamic limit. Thus, $\mathrm{NPR}_n\rightarrow1$ and $\mathrm{NPR}_n\rightarrow0$ represent the fully extended and localized limits, respectively. For finite systems, these quantities generally lie between their limiting values, and their system-size dependence is used to distinguish extended from localized states. The simultaneous decrease of the IPR and increase of the NPR therefore provides a consistent signature of spatial delocalization.

To characterize states that exhibit spatial scaling intermediate between the localized and extended limits, we further consider the fractal dimension associated with the second-order participation moment\cite{D2},
\begin{equation}
D_{2,n}=
-\frac{\ln(\mathrm{IPR}_n)}{\ln L}.
\end{equation}
For a conventionally extended state, $\mathrm{IPR}_n\sim L^{-1}$ and hence $D_{2,n}\rightarrow1$ as $L\rightarrow\infty$, whereas a localized state with finite IPR gives $D_{2,n}\rightarrow0$. Values of $D_{2,n}$ between these limiting values indicate anomalous scaling of the wave-function amplitudes and may be associated with critical or multifractal states. Thus, $D_{2,n}$ provides information beyond the binary distinction between localized and extended states by quantifying the scaling of the eigenstate support with system size. 

In addition to the eigenstate-resolved quantities, we calculate their averages over the $L$ eigenstates,
\begin{equation}
\begin{aligned}
\langle \mathrm{IPR}\rangle
&=
\frac{1}{L}\sum_{n=1}^{L}\mathrm{IPR}_n,
\qquad
\langle \mathrm{NPR}\rangle
=
\frac{1}{L}\sum_{n=1}^{L}\mathrm{NPR}_n,\\
\langle D_2\rangle
&=
\frac{1}{L}\sum_{n=1}^{L}D_{2,n}.
\end{aligned}
\end{equation}
These averaged quantities provide a global characterization of the localization properties and complement the eigenstate-resolved analysis by capturing the overall evolution of the spectrum with the model parameters.

\subsection{Transfer-matrix Lyapunov exponent analysis}

As a complementary probe of localization, we employ a transfer-matrix
analysis to obtain an eigenenergy-resolved Lyapunov diagnostic. The use of
transfer matrices and Lyapunov exponents to characterize localization in
one-dimensional quasiperiodic systems is well established
\cite{lya1,lya2,lya3}, and related formulations have also
been applied to non-Hermitian quasiperiodic systems
\cite{lya5,lya4}. For the present model, the single-particle
Schr\"odinger equation at a generally complex energy $E$ reads
\begin{equation}
t\psi_{n-1}+V_n\psi_n+t\psi_{n+1}=E\psi_n ,
\end{equation}
where $t$ denotes the fixed nearest-neighbor hopping amplitude. Introducing
the two-component vector
$\left(\psi_{n+1},\psi_n\right)^{T}$, the recurrence relation can be expressed
as
\begin{equation}
\begin{pmatrix}
\psi_{n+1}\\
\psi_n
\end{pmatrix}
=
T_n(E)
\begin{pmatrix}
\psi_n\\
\psi_{n-1}
\end{pmatrix},
\end{equation}
where the local transfer matrix is
\begin{equation}
T_n(E)=
\begin{pmatrix}
\dfrac{E-V_n}{t} & -1\\[5pt]
1 & 0
\end{pmatrix}.
\end{equation}
For the Hamiltonian considered here, the onsite term entering this
recurrence is
\begin{equation}
V_n=
W\cos(2\pi b n)
+i\lambda(-1)^n\sin(2\pi b n),
\end{equation}
so that $T_n(E)$ is generally complex even though the hopping amplitude is
real.

For a periodic ring containing $L$ sites, the transfer evolution is carried
out over all $L$ sites. In particular, the closing bond of the ring is
included consistently in the final transfer step, so that the complete
finite-size transfer product can be written as
\begin{equation}
\mathcal{T}(E)
=
T_{L-1}(E)\cdots T_1(E)T_0(E).
\end{equation}
Thus, the indices $n=0,1,\ldots,L-1$ represent $L$ transfer steps rather
than $L-1$ steps. Direct multiplication of the transfer matrices can,
however, lead to numerical overflow or loss of precision. We therefore
stabilize the transfer evolution by performing a QR decomposition after
each step,
\begin{equation}
T_n(E)Q_{n-1}=Q_nR_n,
\qquad Q_0=I,
\end{equation}
where $Q_n$ is orthogonal and $R_n$ is upper triangular. The corresponding
finite-size growth rates are obtained from the diagonal elements of the
successive $R_n$ matrices,
\begin{equation}
\gamma_{\alpha}(E)
=
\frac{1}{L}
\sum_{n=0}^{L-1}
\ln\left|R_n^{\alpha\alpha}\right|,
\qquad \alpha=1,2 .
\end{equation}
We use the largest real part of these two growth rates as the Lyapunov
diagnostic,
\begin{equation}
\boxed{
\gamma{(E)}
=
\max_{\alpha=1,2}
\left\{
\operatorname{Re}
\left[\gamma_{\alpha}(E)\right]
\right\}.
}
\end{equation}

A high value of $\gamma(E)$ indicates a substantial exponential growth or attenuation of the transfer amplitudes and is consistent with localized behavior. Conversely, a suppression of $\gamma(E)$ toward zero indicates
the reduction of exponential transfer-matrix growth and is consistent with extended or critical behavior in the large-system limit. Since the present calculation is performed for finite periodic rings, $\gamma(E)$ need not vanish exactly even for states that are not exponentially localized.
Accordingly, we use the magnitude and systematic suppression of $\gamma(E)$ as a complementary indicator of changes in the spatial character of the eigenstates rather than as a standalone localization criterion.

We evaluate the transfer-matrix Lyapunov diagnostic at the eigen-energies of the non-Hermitian Hamiltonian,
The eigenstates are ordered according to increasing real energy,
which defines the eigenstate index used in the energy-resolved plots.
For each eigenenergy, the full complex value $E_n$ is retained in the
transfer-matrix calculation, and the resulting quantity is denoted by $\gamma(E_n)$.
Thus, $\gamma(E_n)$ provides a state-resolved characterization of the
localization properties across the complex spectrum, rather than an
energy-averaged measure. At fixed $\lambda$, we evaluate
$\gamma(E_n)$ as a function of $W$, while at fixed $W$ we evaluate it as
a function of $\lambda$. Suppression of $\gamma(E_n)$ within a restricted
eigenstate sector, together with the corresponding decrease in IPR and
increase in NPR and $D_2$, provides complementary evidence for a change
in the localization character of that sector.
\subsection{Wave-packet dynamics and RMS spreading}

To provide a dynamical counterpart to the eigenstate-based localization
measures, we examine the real-space evolution of an initially localized
wave packet. The time evolution is generated by the non-Hermitian
Hamiltonian according to
\begin{equation}
|\psi(t)\rangle=e^{-iHt}|\psi(0)\rangle .
\end{equation}
In the non-Hermitian system, the norm of the evolving state is generally
time dependent. We therefore construct the instantaneous spatial
distribution from the normalized site amplitudes,
\begin{equation}
P_n(t)=
\frac{|\psi_n(t)|^2}
{\displaystyle\sum_{j=1}^{L}|\psi_j(t)|^2},
\end{equation}
so that $\sum_n P_n(t)=1$ at every time. This normalization allows the
spatial spreading of the wave packet to be followed independently of the
overall amplification or attenuation generated by the non-Hermitian
dynamics.

The spatial extent of the wave packet is quantified by its root-mean-square
(RMS) width~\cite{dyna1,dyna2,dyna3},
\begin{equation}
\sigma(t)=
\left[
\sum_{n=1}^{L}(n-n_0)^2P_n(t)
\right]^{1/2},
\end{equation}
where $n_0$ is the initial position of the packet. A wave packet that
remains confined to a limited region exhibits a relatively small RMS
width, whereas sustained spreading over the lattice produces a larger
value of $\sigma(t)$. To characterize the overall spreading during the
evolution, we consider the maximum RMS width reached within the chosen
time interval,
\begin{equation}
\sigma_{\max}=\max_t \sigma(t).
\end{equation}
For comparison between different parameter values, this quantity is
normalized by the global maximum obtained over the parameter space, $\sigma_{\max}/\sigma_{\max}^{\mathrm{global}}$.
This dimensionless measure highlights relative changes in the dynamical
spreading while removing the overall scale of $\sigma_{\max}$.

The normalized RMS spreading provides a real-space diagnostic that is
complementary to the static eigenstate measures. Regions with enhanced
$\sigma_{\max}/\sigma_{\max}^{\mathrm{global}}$ correspond to stronger
wave-packet spreading and therefore reduced spatial confinement, whereas
suppressed values indicate stronger dynamical confinement. In particular,
non-monotonic variations of this quantity with $W$ or $\lambda$ can reveal
dynamical signatures of the re-entrant behavior identified from the IPR,
NPR, and $D_2$.

\section{Results and discussion}

In this section, we present the main numerical results and discuss the physical origin of the observed localization behavior. Throughout the calculations, the hopping amplitude is fixed at $t=1$~eV, and all energies are expressed in units of eV. Unless otherwise specified, we consider a periodic ring consisting of $L=610$ lattice sites, while the finite-size analysis is extended up to $L=1597$ sites. We choose Fibonacci numbers for the system sizes to construct periodic approximants of the underlying quasiperiodic modulation, thereby providing a systematic approach to investigating the thermodynamic behavior while minimizing finite-size commensurability effects. The values of the remaining model parameters are specified at the corresponding points throughout the following discussion.
\subsection{Eigenstate resolved localization characteristics}
\begin{figure*}[t]
    \centering
  \includegraphics[width=17cm,height=8cm]{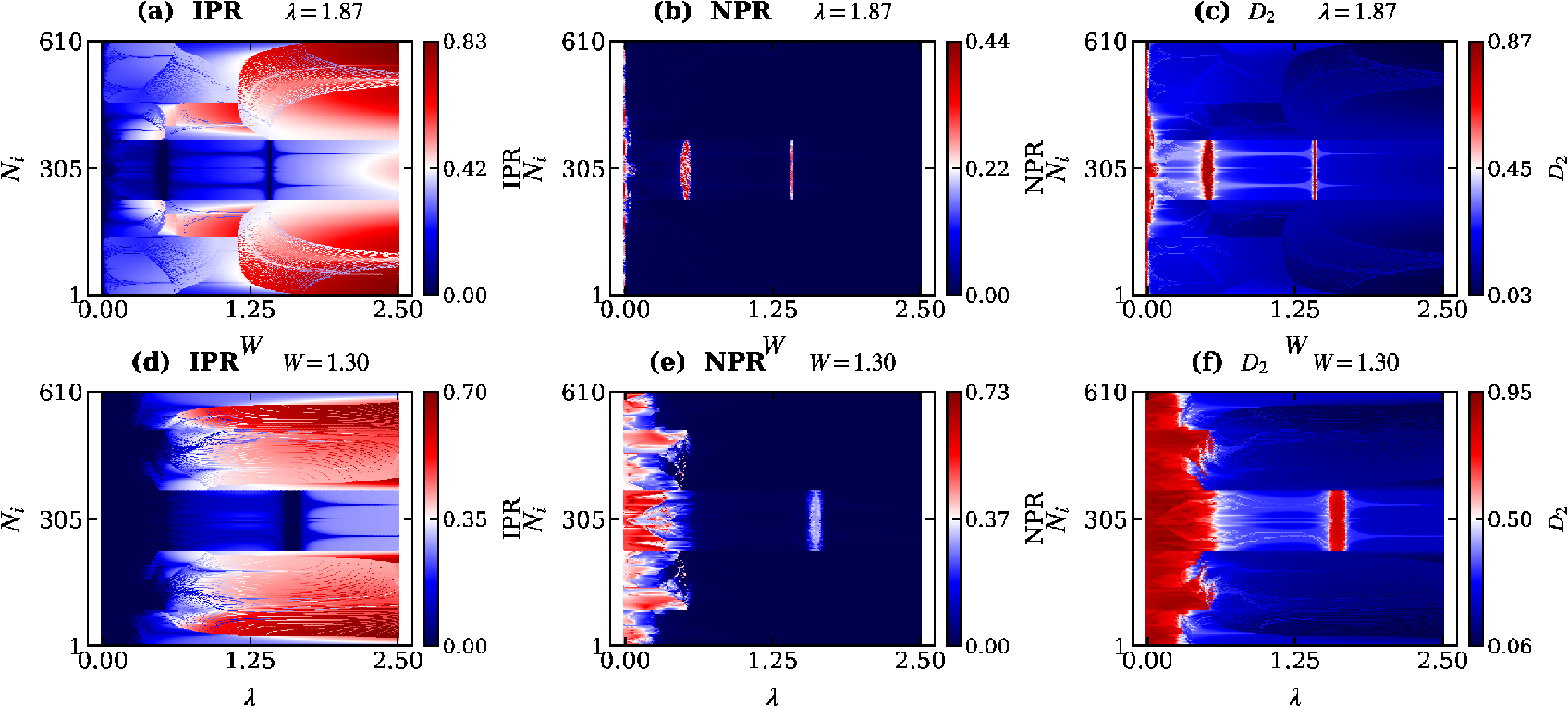}
    \caption{State-resolved localization characteristics as functions of the quasiperiodic potential strength $W$ and the non-Hermitian modulation strength $\lambda$. Panels (a), (b), and (c) show the inverse participation ratio (IPR), normalized participation ratio (NPR), and fractal dimension $D_2$, respectively, as functions of $W$ at fixed $\lambda=1.87$. Panels (d), (e), and (f) display the corresponding quantities as functions of $\lambda$ at fixed $W=1.30$. The color maps resolve the localization properties across the eigenstate index $N_i$, with an independent color scale for each panel.}
    \label{fig:egnstatewlam}
\end{figure*}
Figure~\ref{fig:egnstatewlam} presents the eigenstate-resolved localization
properties in terms of the inverse participation ratio (IPR), normalized
participation ratio (NPR), and fractal dimension $D_2$. The upper panels
Figs.~\ref{fig:egnstatewlam}(a)-(c) show their dependence on the
quasiperiodic potential strength $W$ at fixed non-Hermitian modulation
$\lambda=1.87$, while Figs.~\ref{fig:egnstatewlam}(d)-(f) display the
corresponding evolution with $\lambda$ at fixed $W=1.30$. The eigenstate
index $N_i$ is used to resolve the localization properties across the
spectrum, thereby revealing spectral variations that may be obscured by
averaged quantities.

For the $W$-dependent evolution, a substantial portion of the spectrum
exhibits relatively small IPR and enhanced NPR and $D_2$ in the
weak-potential regime, consistent with extended or weakly localized
eigenstates. Increasing $W$ generally drives the spectrum towards
localization, although this trend is markedly non-monotonic within the
central spectral sector $N_i\sim240$-$375$. In particular, around
$W\approx0.52$, this sector exhibits suppressed IPR and enhanced NPR and
$D_2$, indicating a recovery of extended or critical spatial
character. Upon further increasing $W$, the same eigenstates become more
localized before exhibiting a second delocalization-like response around
$W\approx1.42$. Beyond this region, the localization tendency is
re-established. This sequence of
extended-localized-extended-localized behaviour constitutes a
re-entrant localization response, which is predominantly confined to
a restricted part of the spectrum rather than shared uniformly by all
eigenstates. The corresponding variations in NPR and $D_2$ are consistent
with the IPR distribution, providing complementary information about
the spatial structure of the re-entrant states.

The role of non-Hermitian modulation is further examined by varying
$\lambda$ at fixed $W=1.30$. For $\lambda\lesssim0.65$, a substantial
fraction of the spectrum retains relatively extended character, as
indicated by small IPR and enhanced NPR and $D_2$. With increasing
$\lambda$, the overall tendency is towards localization; however, a
distinct non-monotonic feature emerges near $\lambda\approx1.87$.
Specifically, the central sector $N_i\sim240$-$375$ develops a region
of suppressed IPR accompanied by enhanced NPR and $D_2$, embedded within
a predominantly localized spectral background. Away from this parameter
region, the localization indicators recover their localized-state
character. Thus, tuning the non-Hermitian modulation also produces a
state-selective re-entrant response, demonstrating that the localization
properties are sensitive not only to the control parameters but also
to the position of the eigenstates within the spectrum.

Overall, the state-resolved results reveal a non-monotonic localization
scenario governed by the interplay between quasiperiodicity and
non-Hermiticity. The coexistence of eigenstates with distinct spatial
characteristics at the same parameter values suggests a
mobility-edge-like structure, while the recovery of extended or critical
character within otherwise localized spectral regions highlights the
re-entrant nature of the localization behaviour. These observations
provide a spectral-level perspective for the non-monotonic features
investigated further through averaged localization measures and
dynamical diagnostics.
\subsection{Eigenvalue spectrum with eigenstate-resolved localization}
\begin{figure}[t]
    \centering
  \includegraphics[width=8.5cm,height=6cm]{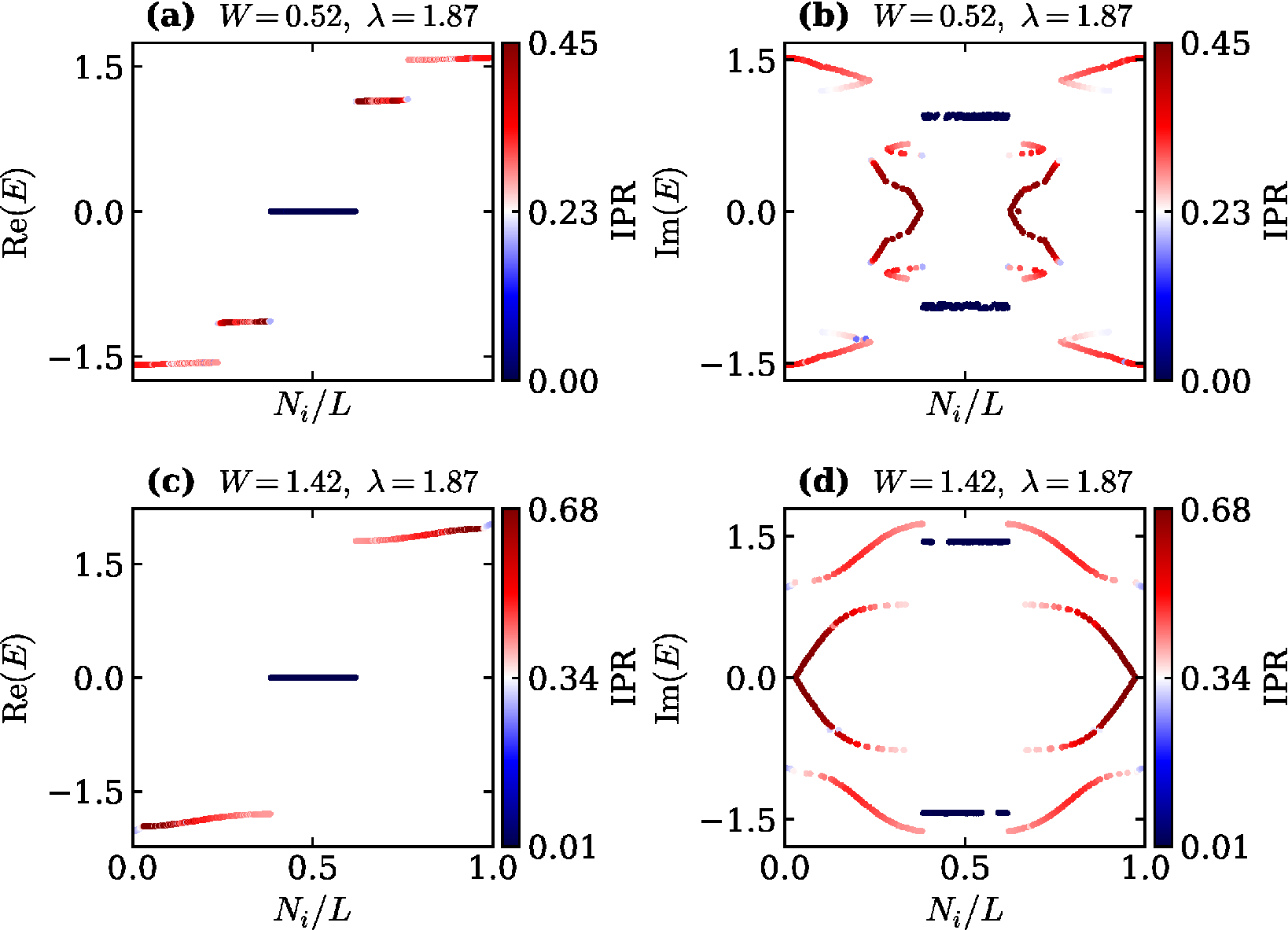}
    \caption{Spatially resolved eigenstate distributions in the complex-energy spectrum.
The first and second rows correspond to $W=0.52$ and $W=1.42$,
respectively, with the non-Hermitian modulation strength fixed at
$\lambda=1.87$. Panels (a) and (b) show $\mathrm{Re}(E)$ and
$\mathrm{Im}(E)$, respectively, as functions of the normalized eigenstate
index $N_i/L$, while panels (c) and (d) show the corresponding quantities
for $W=1.42$. The color scale represents the inverse participation ratio
(IPR), resolving the localization characteristics across the complex
spectrum.}
    \label{fig:reimband}
\end{figure}
To clarify the eigenstate-resolved origin of the re-entrant behavior revealed by the averaged localization measures, we investigate the complex-energy spectrum at two representative quasiperiodic strengths, $W=0.42$ and $W=1.42$. Figure~\ref{fig:reimband} shows the real and imaginary parts of the eigenvalues, with each state color coded by its IPR. The horizontal axis is the normalized eigenstate index $n/L$, obtained by ordering the eigenvalues according to their real parts; it therefore serves only as an eigenstate label and has no momentum-space interpretation.

At $W=0.42$, the real-energy spectrum in Fig.~\ref{fig:reimband}(a) separates into a central branch around $\mathrm{Re}(E)\simeq0$ and upper and lower branches extending toward $\mathrm{Re}(E)\simeq\pm1.5$. The central branch, occupying approximately $n/L\sim0.4$-$0.6$, is predominantly blue, indicating low-IPR and hence spatially extended eigenstates, whereas the outer branches are mainly red and correspond to localized states. The imaginary spectrum [Fig.~\ref{fig:reimband}(b)] resolves the same eigenstate sector into two nearly horizontal blue branches at positive and negative $\mathrm{Im}(E)$. The surrounding curved branches remain predominantly red. Thus, the fragmented complex spectrum separates eigenstates with markedly different localization properties, while the low-IPR states remain confined to a restricted interval of $n/L$.

The same eigenstate-selective organization reappears at $W=1.42$ [Figs.~\ref{fig:reimband}(c) and \ref{fig:reimband}(d)]. A predominantly blue central branch around $\mathrm{Re}(E)\simeq0$ again occurs for $n/L\sim0.4$-$0.6$, whereas the remaining real-energy branches are largely red. In the imaginary-energy plane, this central sector again forms two nearly horizontal blue branches at positive and negative $\mathrm{Im}(E)$, surrounded by predominantly red curved branches. Although the detailed geometry of the complex spectrum changes with $W$, the localization-resolved organization remains essentially unchanged. The recurrence of the same low-IPR eigenstate sector at the two separated values of $W$ therefore provides a direct spectral signature of the two re-entrant delocalization windows.

The fragmented branch structure is characteristic of the complex spectrum of the non-Hermitian quasiperiodic system and, in the present representation, reflects the separation of eigenstates into distinct real- and imaginary-energy sectors. More importantly, the simultaneous presence of low-IPR and high-IPR states at the same parameter values demonstrates that the re-entrant delocalization is strongly state selective rather than a uniform spectral transition. This coexistence of extended and localized eigenstates is consistent with the mobility-edge-like behavior inferred from the IPR, NPR, and fractal-dimension analyses, while the complex-energy representation further reveals how these distinct localization sectors are distributed across the spectrum. \textit{Furthermore, we additionally investigate the complex-energy spectra and find that they do not exhibit closed spectral loops within the parameter regimes considered here. Consequently, we do not assign a point-gap winding number to these spectra. A detailed discussion of the spectral structure, the absence of closed spectral loops, and the corresponding implications for point-gap topology is provided in Appendix~\ref{app:complex_spectrum}.}.

\subsection{Average IPR and NPR: signatures of multiple re-entrant delocalization transitions}
\begin{figure}[t]
    \centering
    \includegraphics[width=8.6cm,height=4cm]{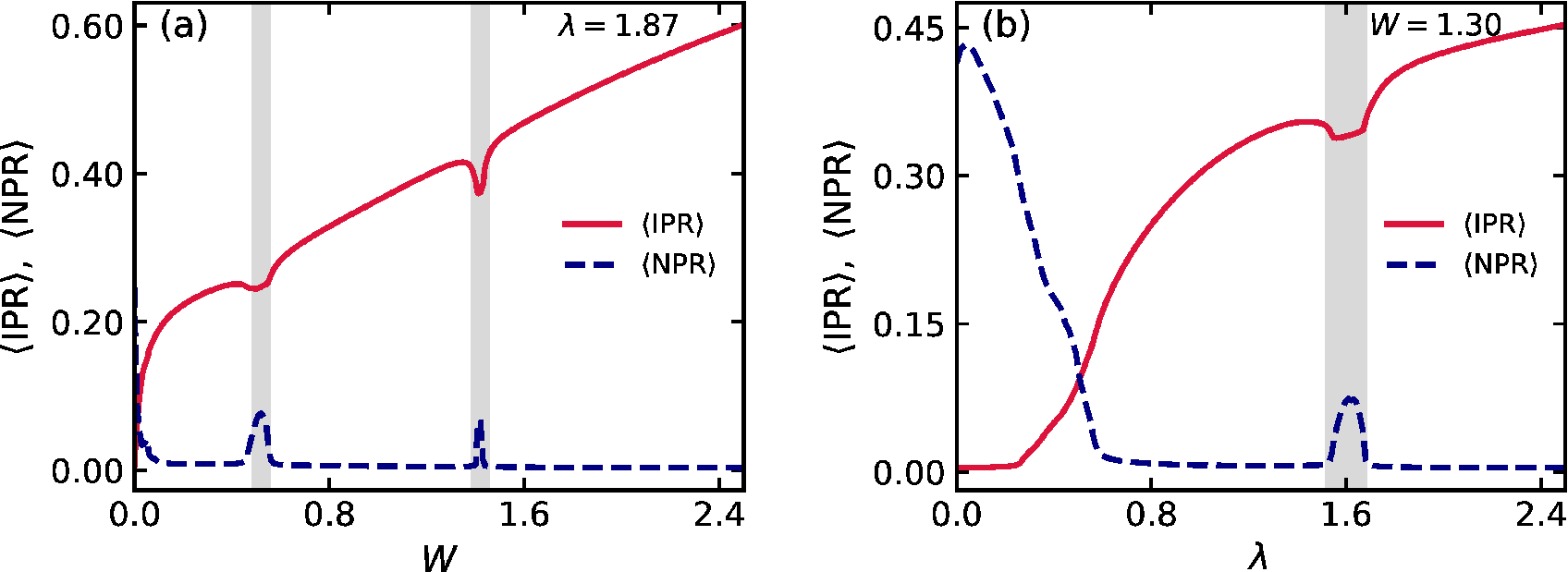}
    \caption{Variation of the $\langle IPR\rangle$ and $\langle NPR\rangle$ with the control parameters. Panel (a) shows the $\langle IPR\rangle$ and $\langle NPR\rangle$ as functions of the quasiperiodic disorder strength $W$ for fix $\lambda=1.87$, while panel (b) presents their dependence on the non-Hermitian modulation $\lambda$ for fix $W=1.3$.}
    \label{fig:avgiprwlam}
\end{figure}
We characterize the overall localization behavior using the $\langle\mathrm{IPR}\rangle$ and $\langle\mathrm{NPR}\rangle$, which provide complementary measures of the spatial extent of the eigenstates. Figure~\ref{fig:avgiprwlam}(a) shows their evolution with the quasiperiodic potential strength $W$ for a fixed non-Hermitian strength $\lambda=1.87$. In the weak-disorder regime, the small value of $\langle\mathrm{IPR}\rangle$ together with a finite $\langle\mathrm{NPR}\rangle$ indicates predominantly extended eigenstates. As $W$ increases, $\langle\mathrm{IPR}\rangle$ generally grows while $\langle\mathrm{NPR}\rangle$ decreases, reflecting an overall tendency toward localization. However, two pronounced non-monotonic features emerge near $W\approx0.52$ and $W\approx1.42$, where noticeable dips in $\langle\mathrm{IPR}\rangle$ are accompanied by corresponding enhancements of $\langle\mathrm{NPR}\rangle$. These features indicate a recovery of spatial extension and suggest the emergence of two re-entrant delocalization windows within the predominantly localized regime. Beyond each window, the increase in $\langle\mathrm{IPR}\rangle$ and suppression of $\langle\mathrm{NPR}\rangle$ signal the restoration of localized behavior. The two separated anomalies therefore reveal a sequence of localization-delocalization-localization crossovers and indicate a non-monotonic localization response arising from the interplay between quasiperiodicity and non-Hermiticity.

The complementary dependence on the non-Hermitian modulation is shown in Fig.~\ref{fig:avgiprwlam}(b), where $\langle\mathrm{IPR}\rangle$ and $\langle\mathrm{NPR}\rangle$ are plotted as functions of $\lambda$ for fixed $W=1.3$. In the weak non-Hermitian regime, the relatively small $\langle\mathrm{IPR}\rangle$ and finite $\langle\mathrm{NPR}\rangle$ indicate predominantly extended eigenstates. With increasing $\lambda$, $\langle\mathrm{NPR}\rangle$ decreases while $\langle\mathrm{IPR}\rangle$ increases, indicating a progressive tendency toward localization. A distinct non-monotonic feature appears near $\lambda\approx1.6$, where a dip in $\langle\mathrm{IPR}\rangle$ is accompanied by a modest enhancement of $\langle\mathrm{NPR}\rangle$. This feature signals a temporary recovery of spatial extension and identifies a re-entrant delocalization window between localized regimes. At larger $\lambda$, $\langle\mathrm{IPR}\rangle$ increases toward saturation while $\langle\mathrm{NPR}\rangle$ remains close to zero, confirming the subsequent restoration of localization. \textit{For comparison, we have also investigated the localization properties of alternative non-Hermitian onsite modulation profiles, as discussed in Appendix~\ref{app:nh_modulation}. Unlike the primary model, these configurations do not exhibit a similar re-entrant localization-delocalization scenario, suggesting that the emergence of multiple re-entrant transitions depends sensitively on the spatial profile of the non-Hermitian onsite modulation.}The consistent non-monotonic response of the two complementary localization measures therefore provides the initial evidence for re-entrant localization-delocalization behavior, which is examined in greater detail through state-resolved, critical, and dynamical analyses below.

\subsection{Correlation between IPR, NPR, and fractal dimension at the two re-entrant delocalization windows}
\begin{figure}[t]
    \centering
   \includegraphics[width=8.5cm,height=4cm]{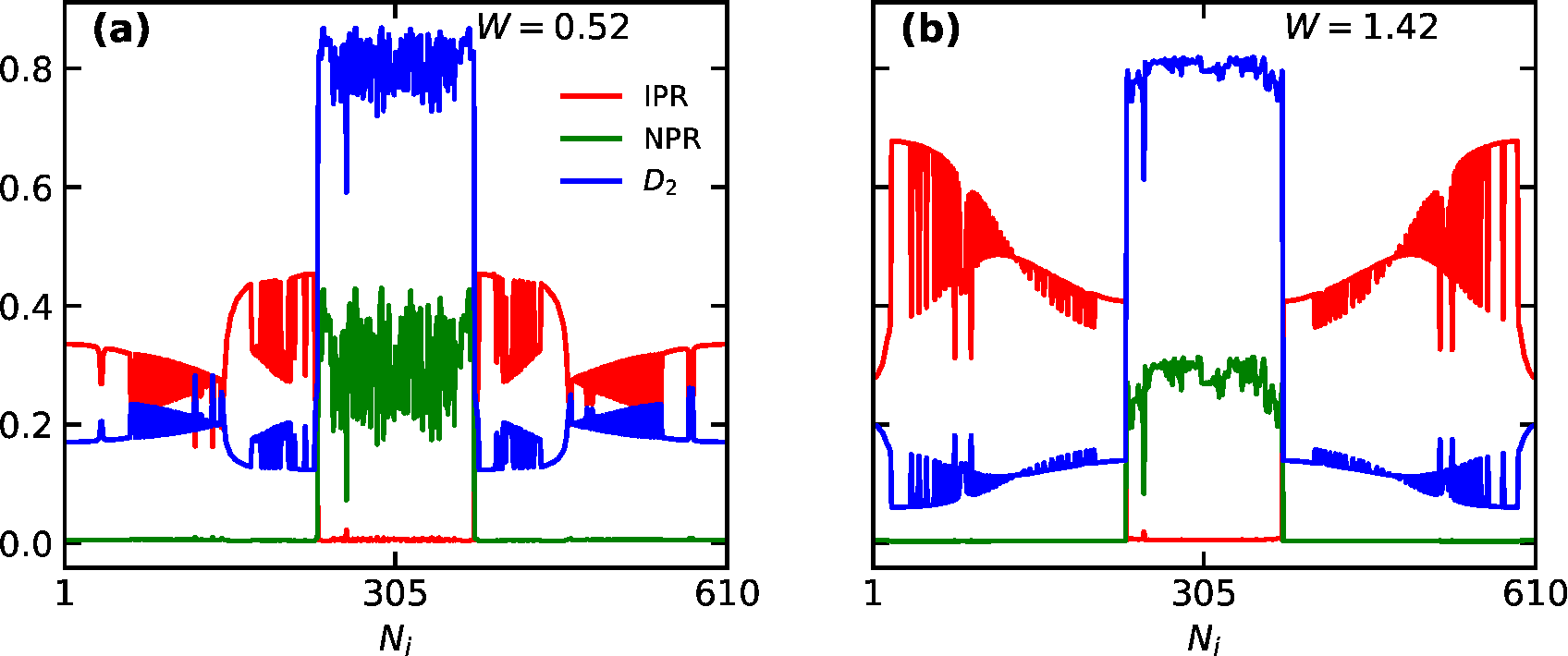}
    \caption{State-resolved localization measures at two representative quasiperiodic
potential strengths. Panels (a) and (b) show the IPR, NPR, and fractal
dimension $D_2$ as functions of the eigenstate index $N_i$ for
$W=0.52$ and $W=1.42$, respectively, at fixed $\lambda=1.87$.}
    \label{fig:iprnprd2}
\end{figure}
\begin{figure}[t]
    \centering
   \includegraphics[width=8.5cm,height=8cm]{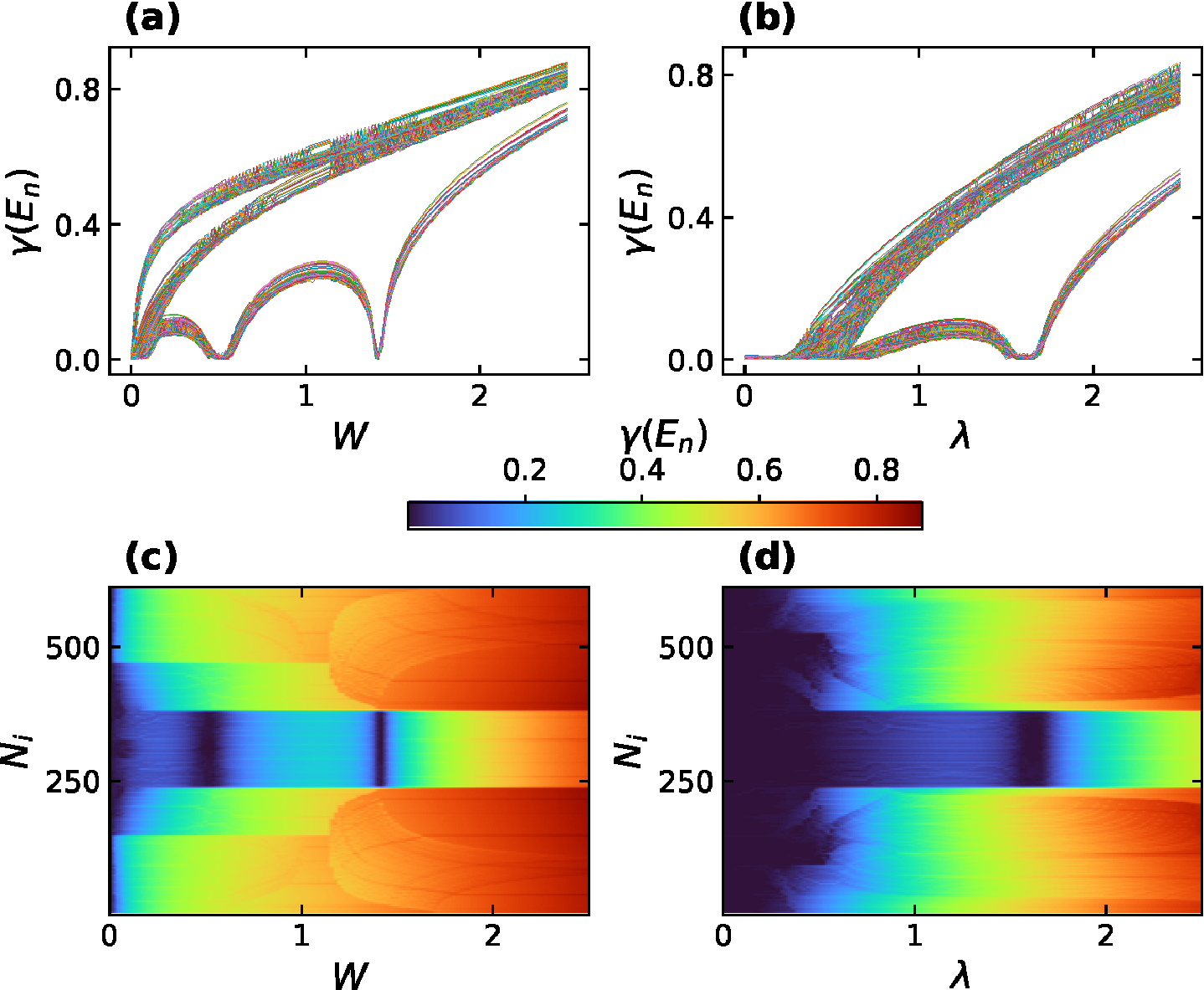}
    \caption{Energy-resolved Lyapunov exponent and its state dependence. Panels (a) and (b) show the Lyapunov exponent $\gamma(E_n)$ as functions of $W$ and $\lambda$, respectively, with the different curves corresponding to the real-energy-sorted eigenstates. Panels (c) and (d) display the corresponding state-resolved Lyapunov exponent in the $(N_i,W)$ and $(N_i,\lambda)$ parameter planes, respectively, revealing the non-monotonic localization response and re-entrant structure.}
    \label{fig:le2}
\end{figure}
\begin{figure*}[t]
    \centering
   \includegraphics[width=1.0\linewidth]{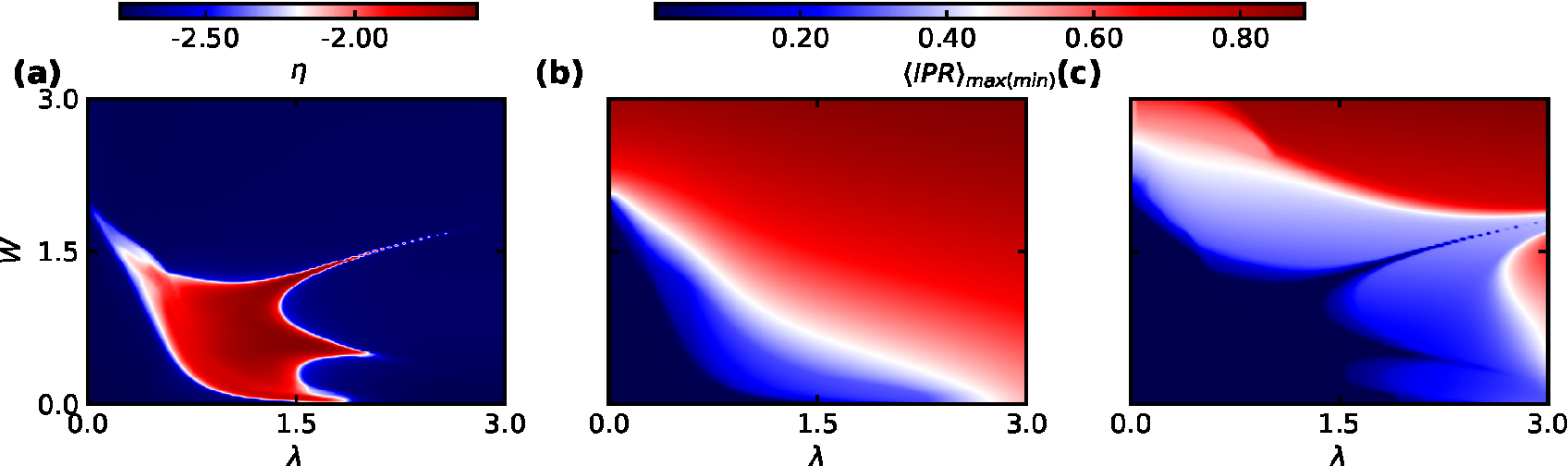}
    \caption{Parameter-dependent localization diagnostics. Panel (a) shows $\eta$ in the
$(\lambda,W)$ parameter plane, while panel (b) presents the corresponding
average $\mathrm{IPR}_{\max}$ over the same parameter space. Panel (c)
displays the average $\mathrm{IPR}_{\min}$ in the $(\lambda,W)$ plane,
highlighting the distinct localization regimes and the boundaries
separating them.}
    \label{fig:eta}
\end{figure*}
To further verify the nature of the re-entrant delocalization windows identified above, we examine the eigenstate-resolved distributions of the IPR, NPR, and fractal dimension $D_2$ at the two representative quasiperiodic potential strengths, $W=0.52$ and $W=1.42$. These values correspond to the two distinct delocalization windows identified from the earlier analyses. Figure~\ref{fig:iprnprd2} displays the three localization measures as functions of the eigenstate index $N_i$, allowing the localization character of the individual eigenstates participating in the two re-entrant regimes to be resolved directly.

Figure~\ref{fig:iprnprd2}(a) corresponds to the first re-entrant delocalization window at $W=0.52$. A well-defined group of eigenstates near the center of the spectrum, approximately $N_i\approx230$-$380$, exhibits a pronounced suppression of the IPR accompanied by an enhancement of both the NPR and $D_2$. Thus, the three measures show a consistent correspondence: eigenstates with reduced IPR simultaneously acquire larger NPR and $D_2$, indicating enhanced spatial extension and, for intermediate values, critical character. Away from this central eigenstate sector, the IPR increases while the NPR and $D_2$ are suppressed, showing that a substantial part of the remaining spectrum retains localized character. The same correlated behavior reappears in Fig.~\ref{fig:iprnprd2}(b) at the second re-entrant window, $W=1.42$. The corresponding central group of eigenstates again shows low IPR together with enhanced NPR and $D_2$, whereas the states outside this sector remain predominantly localized.

The recurrence of the same eigenstate-selective correspondence at two well-separated values of $W$ provides direct evidence that the re-entrant delocalization is a robust feature of the eigenstate structure rather than an isolated numerical effect. In particular, the simultaneous enhancement of NPR and $D_2$ with the suppression of IPR demonstrates that the two re-entrant windows are associated with genuine changes in the spatial character of the wave functions. At the same time, the persistence of localized states outside the selected eigenstate sector shows that the delocalization is not uniform across the spectrum. This coexistence of localized and extended or critical states at the same parameter values further supports the mobility-edge-like character of the localization transitions in the non-Hermitian quasiperiodic system.

\subsection{Energy-resolved Lyapunov exponent and re-entrant localization}

To further elucidate the state-resolved character of the re-entrant localization-delocalization transitions, we investigate the energy-dependent Lyapunov exponent $\gamma(E_n)$. Here, the Lyapunov exponent is evaluated at the eigenvalues $E_n$ of the corresponding Hamiltonian, providing a localization measure for individual eigenstates rather than an averaged characterization of the spectrum. This analysis allows us to determine whether the re-entrant behavior identified above is distributed uniformly across the spectrum or confined to specific eigenstate sectors. Figure~\ref{fig:le2} presents the resulting $\gamma(E_n)$ as functions of $W$ and $\lambda$.

In Fig.~\ref{fig:le2}(a), the $\gamma(E_n)$ curves exhibit a distinctly non-monotonic dependence on $W$. For a subset of the real-energy-sorted eigenstates, $\gamma(E_n)$ is strongly suppressed around $W\simeq0.52$ and develops a second pronounced minimum near $W\simeq1.42$. Since a finite positive Lyapunov exponent indicates exponential localization, whereas $\gamma(E_n)\rightarrow0$ signals the disappearance of exponential localization, these two minima identify two distinct parameter regions in which the corresponding eigenstates become substantially less localized. Their occurrence at two separated values of $W$ therefore provides an eigenstate-resolved signature of the two re-entrant delocalization windows identified above. A related behavior is observed in Fig.~\ref{fig:le2}(b), where selected eigenstate branches exhibit a pronounced suppression of $\gamma(E_n)$ around $\lambda\simeq1.87$, indicating that the delocalized response can likewise be recovered by tuning the non-Hermitian modulation.

The eigenstate-resolved maps in Figs.~\ref{fig:le2}(c) and (d) further identify the portions of the spectrum involved in these transitions. In both maps, the reduced-$\gamma$ regions form well-defined bands over restricted ranges of the real-energy-sorted eigenstate index $N_i$, rather than extending uniformly across the complete spectrum. In particular, the low-$\gamma$ bands in Fig.~\ref{fig:le2}(c) correspond to the two parameter regions around $W\simeq0.52$ and $W\simeq1.42$ where the branches in Fig.~\ref{fig:le2}(a) develop pronounced minima, while the low-$\gamma$ structure in Fig.~\ref{fig:le2}(d) is consistent with the suppression observed near $\lambda\simeq1.87$ in Fig.~\ref{fig:le2}(b). Thus, the Lyapunov-exponent analysis independently confirms that the re-entrant delocalization is concentrated in specific portions of the eigenstate spectrum rather than occurring uniformly across all states. This state selectivity, together with the coexistence of localized and delocalized states identified from the earlier analyses, supports a mobility-edge-like localization scenario in the non-Hermitian quasiperiodic system.

\subsection{Phase diagram from spectral and localization Diagnostics}

To further clarify the microscopic origin of the re-entrant localization-delocalization behavior and its dependence on individual eigenstates, we examine the localization landscape over the full $(\lambda,W)$ parameter space. For this purpose, Fig.~\ref{fig:eta} combines the spectral quantity $\eta$ with the extremal eigenstate-resolved localization measures $\langle\mathrm{IPR}_{\max}\rangle$ and $\langle\mathrm{IPR}_{\min}\rangle$. The $\eta$ map in Fig.~\ref{fig:eta}(a) reveals the overall phase structure, while the two IPR measures provide a direct spatial characterization of the corresponding eigenstates. In Fig.~\ref{fig:eta}(b), the blue region, where $\langle\mathrm{IPR}_{\max}\rangle$ approaches zero, indicates that even the eigenstates with the largest IPR are predominantly extended, whereas the red region signifies the presence of strongly localized states; the blue-white region represents an intermediate regime. The complementary $\langle\mathrm{IPR}_{\min}\rangle$ map in Fig.~\ref{fig:eta}(c) shows the opposite limiting behavior: the red region, with large $\langle\mathrm{IPR}_{\min}\rangle$, indicates that all eigenstates remain localized, while the blue region, where $\langle\mathrm{IPR}_{\min}\rangle$ is small, indicates the presence of extended eigenstates, with the blue-white region corresponding to an intermediate state. Consequently, the region characterized by large $\langle\mathrm{IPR}_{\max}\rangle$ together with small $\langle\mathrm{IPR}_{\min}\rangle$ identifies a spectrum containing both localized and extended eigenstates, providing direct evidence of state-dependent localization and mobility-edge-like behavior. The close correspondence between these IPR-based structures and the boundaries identified from $\eta$ and the critical-transition analysis establishes that the re-entrant behavior arises from genuine changes in the spatial character of the eigenstates across the $(\lambda,W)$ parameter space.

\subsection{Critical parameters and transition boundaries}
\begin{figure}[t]
    \centering
   \includegraphics[width=8.5cm,height=4.5cm]{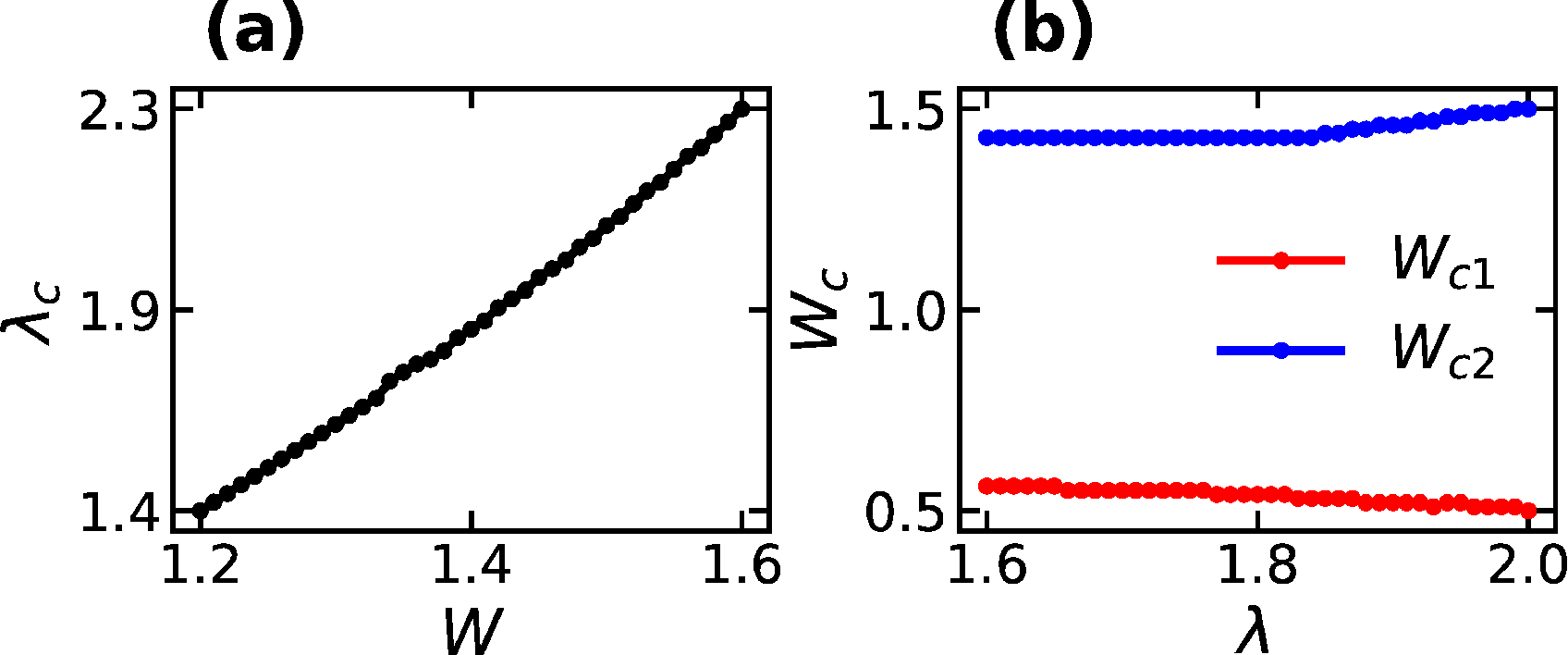}
    \caption{Critical parameters characterizing the localization transitions. Panel (a) shows the critical quantity $\lambda_c$ as a function of the quasiperiodic disorder strength $W$. Panel (b) displays the two critical disorder strengths, $W_{c1}$ and $W_{c2}$, as functions of the non-Hermitian modulation $\lambda$, revealing two distinct transition boundaries and an intermediate parameter regime.}
    \label{fig:lamcwc}
\end{figure}
The robustness of the re-entrant delocalization windows is further examined by tracking their corresponding critical boundaries in the $(W,\lambda)$ parameter space. Figure~\ref{fig:lamcwc} shows the evolution of these critical boundaries with the control parameters. In Fig.~\ref{fig:lamcwc}(a), $\lambda_c$ increases smoothly with $W$, indicating a systematic shift of the critical condition as the quasiperiodic potential is strengthened. The absence of abrupt features in the $\lambda_c$-$W$ relation suggests that the transition boundary evolves continuously rather than through a sudden change in its character. Thus, the curve traces the critical boundary separating the corresponding localization and re-entrant delocalization regimes as the disorder strength is varied.

A distinct feature appears in Fig.~\ref{fig:lamcwc}(b), where two critical boundaries, $W_{c1}$ and $W_{c2}$, emerge as functions of $\lambda$. The lower boundary $W_{c1}$ shows a decreasing trend with increasing $\lambda$, whereas the upper boundary $W_{c2}$ exhibits a more pronounced upward shift at larger $\lambda$. The persistent separation between $W_{c1}$ and $W_{c2}$ therefore defines an intermediate region bounded by two distinct transition lines, demonstrating that the system does not undergo a single direct localization transition. Instead, the two boundaries delimit successive localization and delocalization regimes, consistent with the re-entrant behavior identified from the earlier analysis. The different $\lambda$ dependence of $W_{c1}$ and $W_{c2}$ further indicates that the two transition boundaries respond differently to the non-Hermitian modulation.

\subsection{Finite-size scaling of the average localization measures}
\begin{figure}[t]
    \centering
   \includegraphics[width=8.5cm,height=4cm]{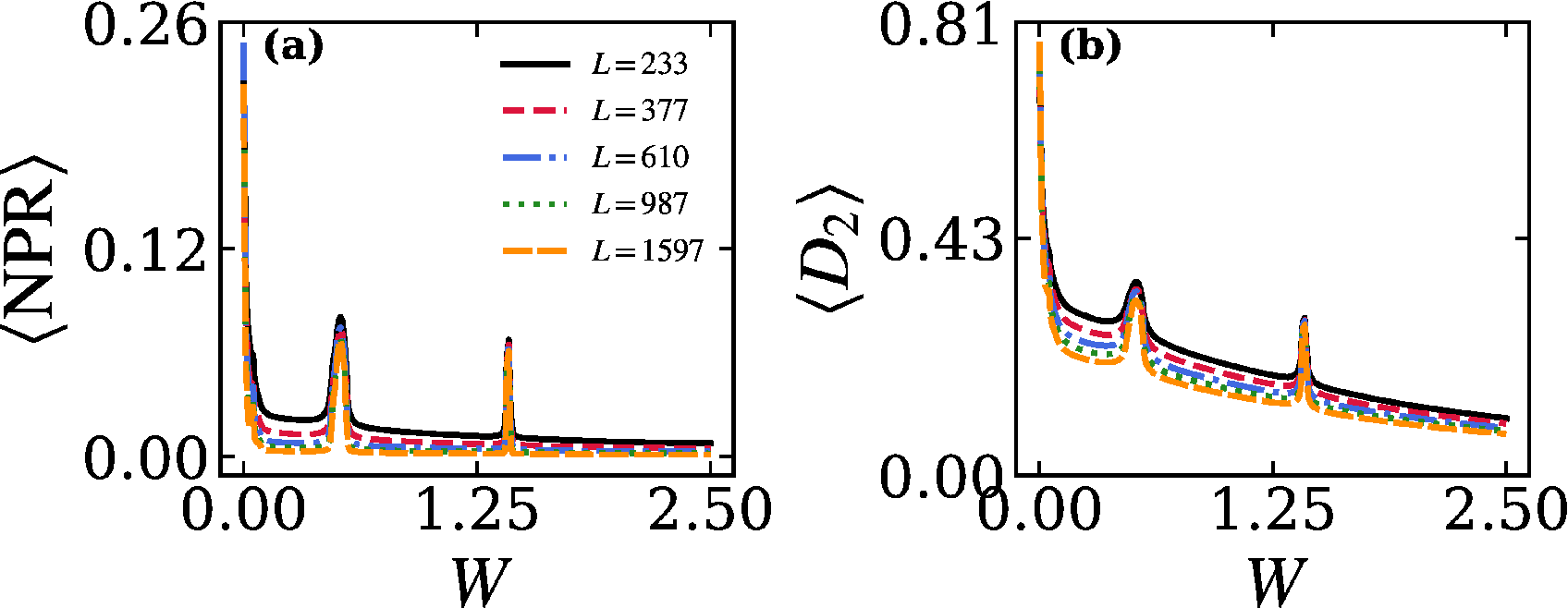}
    \caption{Finite-size scaling of the $\langle\mathrm{NPR}\rangle$ and
$\langle D_2\rangle$. Panel (a) shows $\langle\mathrm{NPR}\rangle$ as a function of $W$ for
different system sizes, $L=233$, $377$, $610$, $987$, and $1597$. Panel (b)
shows the corresponding finite-size behavior of $\langle D_2\rangle$.}
    \label{fig:d2nprscaling}
\end{figure}
\begin{figure}
    \centering
   \includegraphics[width=8.5cm,height=4cm]{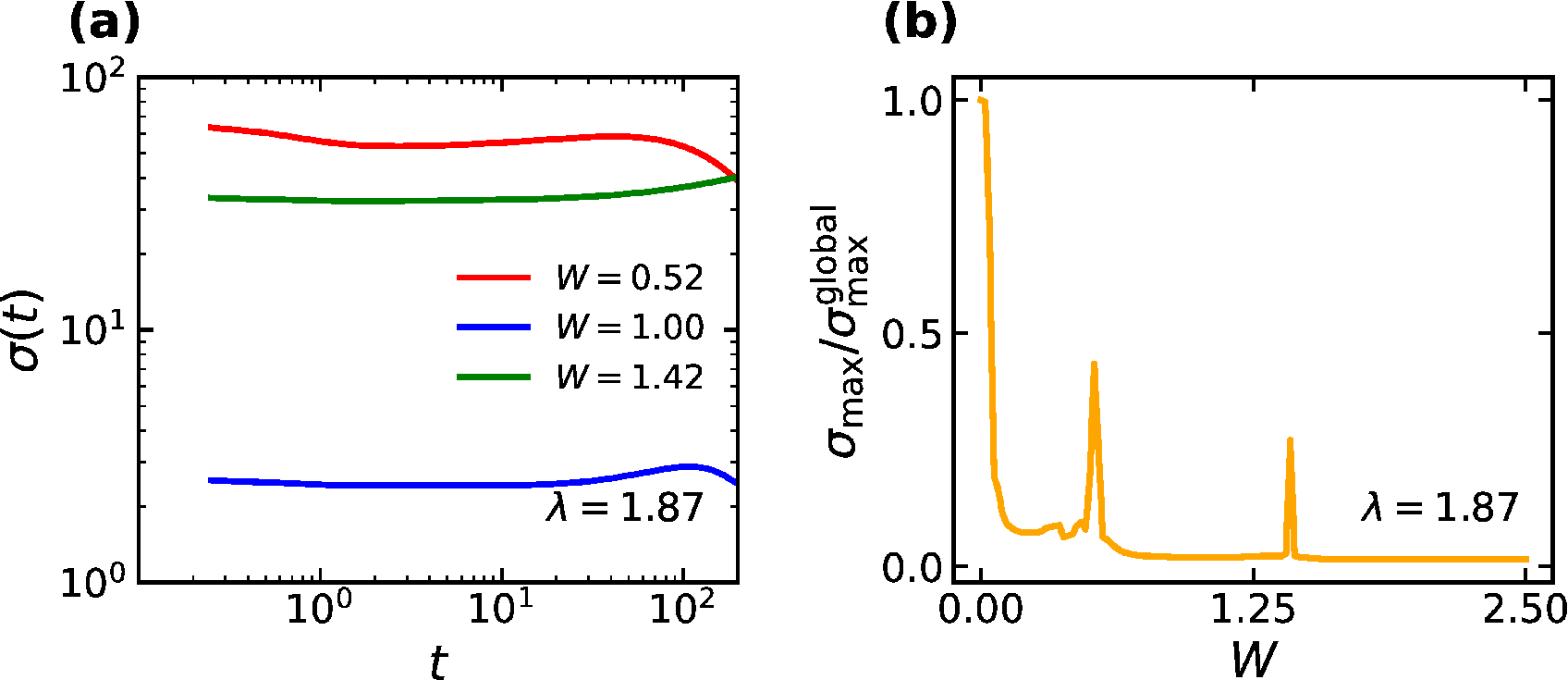}
    \caption{Dynamical signatures of localization transitions through the RMS width. (a) Time evolution of the RMS width, $\sigma(t)$, for three representative values of the quasiperiodic potential strength, $W=0.52$, $1.00$, and $1.42$, at a fixed non-Hermitian modulation strength $\lambda=1.87$. (b) Normalized maximum RMS width, $\sigma_{\max}/\sigma_{\max}^{\mathrm{global}}$, as a function of the quasiperiodic potential strength $W$ at fixed $\lambda=1.87$. The dynamics is obtained by expanding an initially localized state at the central lattice site in the complete right-eigenvector basis and retaining the eigenstates ranked from $240$ to $375$ (inclusive), corresponding to $136$ eigenstates, after sorting the eigenvalues in ascending order of their real parts. The maximum RMS width, $\sigma_{\max}$, is determined over the time interval $0\leq t\leq 200$, and $\sigma_{\max}^{\mathrm{global}}$ denotes the maximum value of $\sigma_{\max}$ over the entire $W$ sweep. Panel (a) is plotted on a log-log scale.}
    \label{fig:RMS}
\end{figure}
To establish that the observed re-entrant localization-delocalization transitions are intrinsic to the system rather than artifacts of finite-size effects, we investigate the system-size dependence of the averaged normalized participation ratio $\langle NPR\rangle$ and fractal dimension $\langle D_2\rangle$. Figure~\ref{fig:d2nprscaling}(a) shows $\langle \mathrm{NPR}\rangle$, while Fig.~\ref{fig:d2nprscaling}(b) presents the corresponding $\langle D_2\rangle$ as functions of the quasiperiodic potential strength $W$ for Fibonacci system sizes ranging from $L=233$ to $L=1597$. For all system sizes considered, both $\langle \mathrm{NPR}\rangle$ and $\langle D_2\rangle$ exhibit two pronounced maxima near $W\approx0.52$ and $W\approx1.42$, coinciding with the two re-entrant delocalization windows identified from the eigenstate-resolved analysis. At these values of $W$, the increase in $\langle \mathrm{NPR}\rangle$ indicates a substantial enhancement of the spatial extent of the eigenstates, while the corresponding peaks in $\langle D_2\rangle$ provide independent evidence of increased fractal support. The consistent two-peak structure of $\langle \mathrm{NPR}\rangle$ and $\langle D_2\rangle$ across system sizes therefore confirms that the observed re-entrant delocalization is robust against finite-size effects.

Away from the two delocalized windows, both $\langle \mathrm{NPR}\rangle$ and $\langle D_2\rangle$ decrease rapidly with increasing $W$, while their overall magnitudes decrease with system size, consistent with the expected scaling of localized states toward the thermodynamic limit. In contrast, the peak positions near $W\approx0.52$ and $W\approx1.42$ remain essentially unchanged, with only moderate reductions in amplitude, demonstrating the robustness of the two delocalized windows against finite-size effects. These results, together with the eigenstate-resolved IPR, NPR, and $D_2$ analysis, establish two robust re-entrant localization-delocalization windows and support the persistence of the corresponding localization-delocalization transitions in the thermodynamic limit.

\subsection{Dynamical signature from the normalized RMS width}

The dynamical evolution of the wave packet provides a complementary probe of the localization properties through the root-mean-square (RMS) width, $\sigma(t)$. Figure~\ref{fig:RMS}(a) shows the time evolution of $\sigma(t)$ for three representative values of the quasiperiodic potential strength, $W=0.52$, $1.00$, and $1.42$, at a fixed non-Hermitian modulation strength $\lambda=1.87$. A clear non-monotonic response is observed as $W$ is varied. For $W=0.52$, the wave packet exhibits a relatively large spatial spread throughout the evolution, whereas at $W=1.00$ its RMS width is strongly suppressed, indicating substantially more confined dynamics. Upon further increasing the potential strength to $W=1.42$, the RMS width increases again, demonstrating a recovery of spatial spreading. Thus, the sequence of enhanced spreading, suppressed spreading, and subsequent recovery provides a direct dynamical signature of re-entrant behavior as a function of $W$.

This non-monotonic behavior is further substantiated by the complete parameter dependence of the normalized maximum RMS width, $\sigma_{\max}/\sigma_{\max}^{\mathrm{global}}$, presented in Fig.~\ref{fig:RMS}(b). Here, $\sigma_{\max}$ is obtained by maximizing $\sigma(t)$ over the evolution interval $0\leq t\leq 200$. After the initial high value near $W=0$, the normalized maximum RMS width decreases sharply, indicating a regime of suppressed wave-packet spreading. Importantly, it subsequently develops a secondary enhancement around $W\simeq 0.52$, followed by another suppression and a pronounced peak around $W\simeq 1.42$. These separated enhancements, interspersed with regions of reduced RMS width, reveal a non-monotonic and re-entrant dynamical response. In particular, the secondary enhancement near $W\simeq 0.52$ and the pronounced peak near $W\simeq 1.42$ are consistent with the enhanced spreading observed for the representative parameter values in Fig.~\ref{fig:RMS}(a). The recovery of the RMS width after an intervening suppressed regime indicates a restoration of enhanced wave-packet spreading at larger potential strength. This dynamical behavior is consistent with the non-monotonic enhancements observed in the static NPR, thereby providing a complementary dynamical signature of the re-entrant localization behavior. \textit{The eigenstate window comprising states $240$-$375$ was selected based on the spectral-resolved Lyapunov-exponent analysis, which identifies this range as an approximate low-Lyapunov-exponent sector associated with enhanced spatial extension and non-monotonic localization characteristics. In particular, this spectral region encompasses the eigenstates relevant to the observed re-entrant localization behavior. This selection also allows us to focus on the spectral sector associated with the non-trivial localization features observed in the static IPR and NPR analyses, while excluding spectral sectors exhibiting comparatively trivial localization behavior. Consequently, restricting the dynamics to this window enables us to investigate the dynamical signatures of the re-entrant localization behavior within the relevant spectral sector.} Accordingly, restricting the dynamics to these $136$ eigenstates provides a targeted investigation of the dynamical response of the relevant spectral sector. The corresponding right eigenvectors are obtained by sorting the eigenvalues in ascending order of their real parts.

\begin{figure*}[t]
    \centering
   \includegraphics[width=12cm,height=10cm]{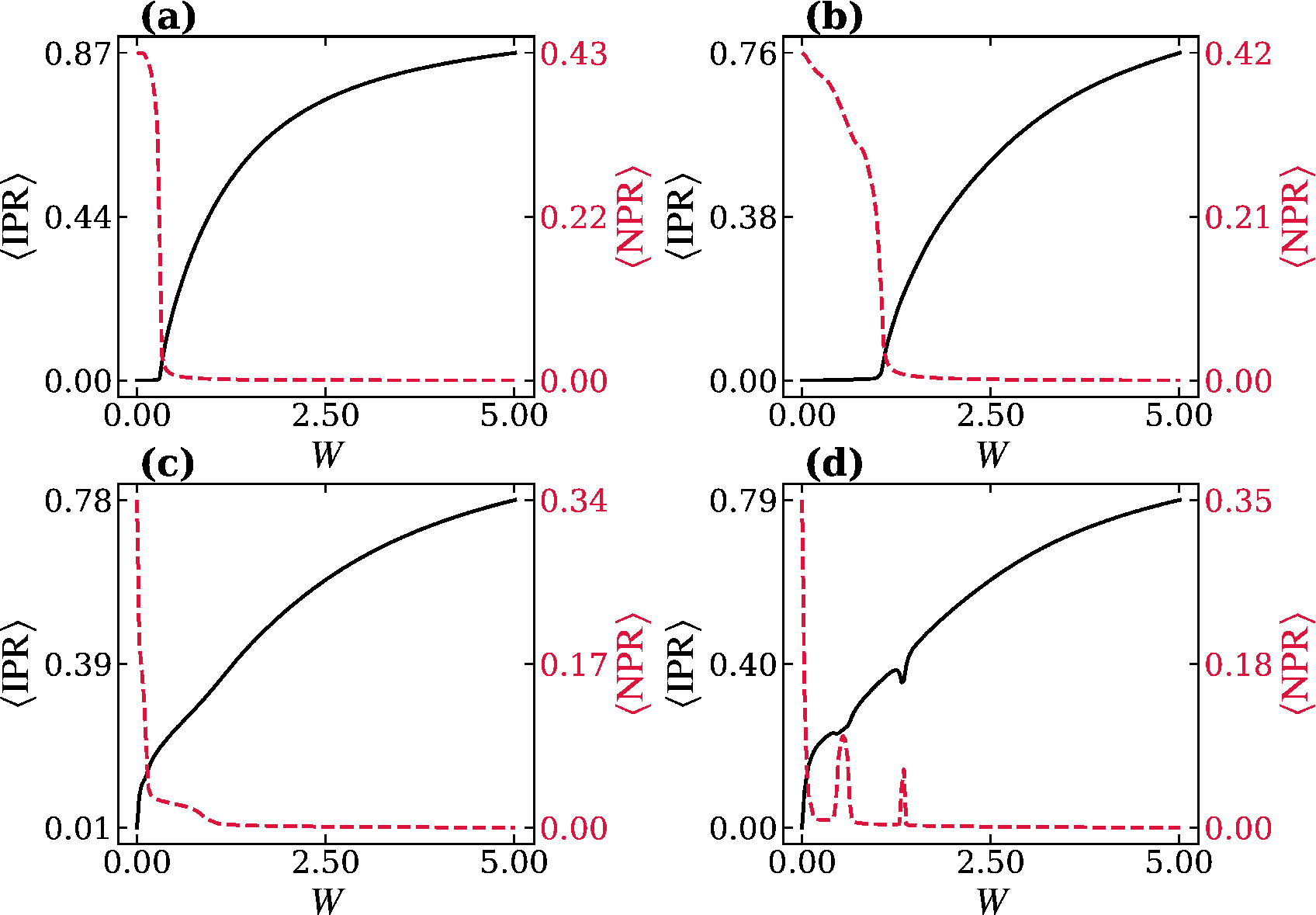}
   \caption{Comparison of the localization response for four spatial forms of the
non-Hermitian onsite modulation at fixed $\lambda=1.7$. The solid black and
dashed red curves represent $\langle\mathrm{IPR}\rangle$ and
$\langle\mathrm{NPR}\rangle$, respectively, as functions of the
quasiperiodic potential strength $W$. Panels (a)-(d) correspond to the
four non-Hermitian modulation forms considered.
}
    \label{fig:modulation_comparison}
\end{figure*}
\section{Conclusion}

We have investigated localization in a dimerized one-dimensional quasiperiodic system with a staggered imaginary quasiperiodic potential, focusing on the combined effects of the quasiperiodic strength $W$ and non-Hermitian strength $\lambda$. The eigenstate-resolved IPR and fractal dimension $D_2$, together with the $\langle IPR\rangle$ and $\langle NPR\rangle$, reveal a distinctly non-monotonic localization response. At fixed $\lambda$, two separated delocalization windows emerge near $W\simeq0.52$ and $W\simeq1.42$, where the suppression of IPR is accompanied by enhanced NPR and $D_2$, followed by a recovery of localized behavior. A similar re-entrant window appears upon varying $\lambda$ at fixed $W$. The corresponding $\eta$, $\mathrm{IPR}_{\max}$, and $\mathrm{IPR}_{\min}$ distributions further establish the existence of an intermediate localization regime. In the central eigenstate sector, extended or critical states coexist with localized states, demonstrating that the transition is strongly state selective rather than a uniform change across the spectrum. The energy-resolved Lyapunov exponent independently identifies the same low-localization sectors through its pronounced suppression at the re-entrant parameter values. Together with the simultaneous presence of localized and extended or critical eigenstates, these results establish a mobility-edge-like localization structure in which the re-entrant response is intrinsically state selective.

The robustness of this structure is supported by its persistence with increasing system size in the averaged NPR and $D_2$, while the associated critical boundaries form two distinct transition lines in the $(W,\lambda)$ parameter space rather than a single monotonic boundary. The complex-energy spectra further reveal the separation of the low-IPR states into restricted spectral branches, including the two positive- and negative-$\mathrm{Im}(E)$ branches associated with the same central eigenstate sector, whereas the surrounding branches remain predominantly localized. Importantly, the static picture is also reproduced dynamically: the normalized RMS wave-packet width develops pronounced enhancements at the parameter values corresponding to the static delocalization windows, providing an independent real-space signature of the re-entrant transitions. Thus, the combined spectral, eigenstate-resolved, scaling, Lyapunov, and dynamical analyses establish a consistent sequence of localization-delocalization-localization regimes generated by the interplay of quasiperiodicity, dimerization, and staggered non-Hermiticity. The resulting state-selective and re-entrant localization structure demonstrates how spatially structured non-Hermitian modulation can substantially reshape both the eigenstate character and the real-space dynamics of quasiperiodic systems.

\appendix

\begin{figure*}[t]
    \centering
   \includegraphics[width=1.0\linewidth]{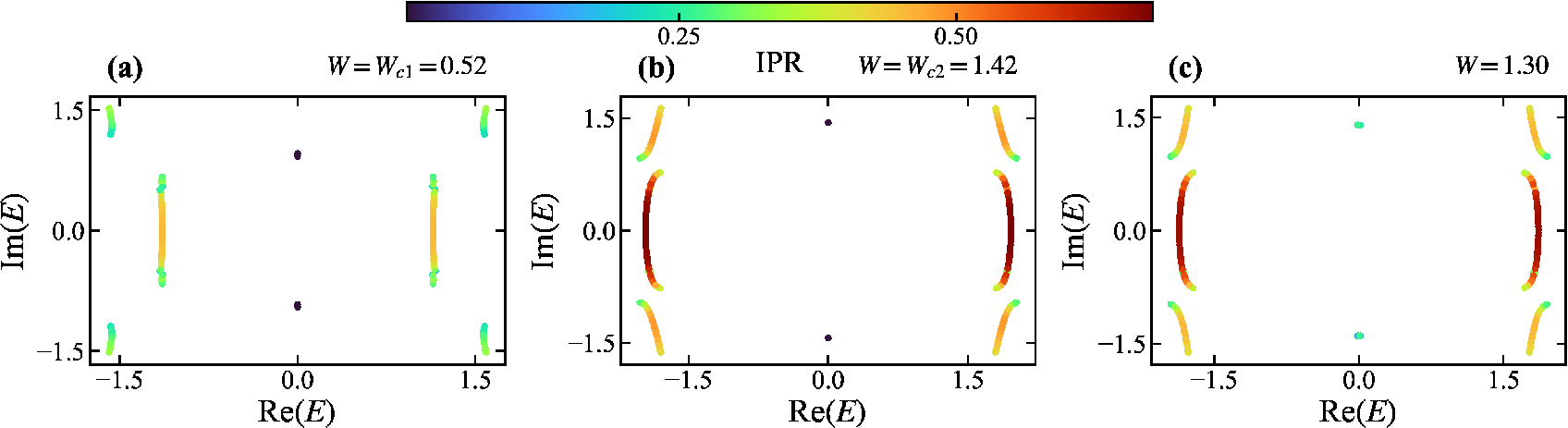}
    \caption{Complex-energy spectra at fixed $\lambda=1.87$ for three representative values of the quasiperiodic disorder strength. The eigenvalues are plotted in the complex-energy plane, with $\mathrm{Re}(E)$ and $\mathrm{Im}(E)$ along the horizontal and vertical axes, respectively, and the color scale representing the IPR. Panels (a)-(c) correspond to $W=W_{c1}=0.52$, $W=W_{c2}=1.42$, and $W=1.30$, respectively.}
    \label{fig:recom}
\end{figure*}

\section{Physical basis for the non-Hermitian onsite modulation}
\label{app:nh_modulation}

To investigate the influence of the spatial profile of the non-Hermitian (NH) onsite modulation on localization, we consider four distinct modulation profiles while keeping the hopping amplitudes, real quasiperiodic potential, and all other model parameters unchanged.
The real part is kept as $W\cos(2\pi b n)$, whereas the imaginary component
is taken successively as
$\mathrm{i}\lambda\sin(2\pi b n)$,
$\mathrm{i}\lambda\cos(2\pi b n)$,
$\mathrm{i}\lambda(-1)^n\cos(2\pi b n)$, and
$\mathrm{i}\lambda(-1)^n\sin(2\pi b n)$.
The last form is used in the main text. For this comparison, we choose $\lambda=1.7$ as a representative value
within the range of $\lambda$ over which the re-entrant response is
observed for the staggered sinusoidal modulation. This value is used
specifically to resolve and compare the localization responses associated
with the different spatial forms of the non-Hermitian potential. In the
main analysis, we instead fix $\lambda=1.87$, which also lies within the
same re-entrant regime and provides a representative parameter point for
the detailed investigation of the localization transitions. Thus, the
choice of $\lambda=1.7$ in this comparison and $\lambda=1.87$ in the main
text reflects two representative parameter values within the re-entrant
range, rather than a qualitative difference between the two cases.
The corresponding results are shown in Fig.~\ref{fig:modulation_comparison}.
For the first two forms (Figs.~\ref{fig:modulation_comparison}(a) and
\ref{fig:modulation_comparison}(b)), $\langle\mathrm{IPR}\rangle$ varies
smoothly with $W$, while $\langle\mathrm{NPR}\rangle$ decreases without
pronounced intermediate enhancement. The staggered cosine form
[Fig.~\ref{fig:modulation_comparison}(c)] modifies this evolution but does
not produce the pronounced re-entrant structure observed for the fourth
form. In contrast, the staggered sinusoidal modulation
[Fig.~\ref{fig:modulation_comparison}(d)] produces two distinct
enhancements of $\langle\mathrm{NPR}\rangle$, accompanied by corresponding
reductions of $\langle\mathrm{IPR}\rangle$, as $W$ is varied. This
non-monotonic response identifies two parameter windows with reduced
spatial localization.

Although the re-entrant response occurs only within certain ranges of
$\lambda$, the comparison shows that it is not obtained for the other
three spatial forms over the parameter regime.
This behavior motivates the choice of the staggered sinusoidal modulation
as the representative NH term in the main text. Its alternating sign
structure, combined with the sinusoidal quasiperiodic dependence shared
with the real onsite potential, provides the spatial competition required
for the pronounced non-monotonic localization response. We therefore fix
this form in the subsequent analysis.
\section{Complex-energy spectrum and absence of a closed spectral loop}
\label{app:complex_spectrum}

The complex-energy spectrum provides a complementary view of the localization properties and the effects of the non-Hermitian onsite modulation. Figure~\ref{fig:recom} displays the eigenvalues in the complex-energy plane for three representative values of $W$ at fixed $\lambda=1.87$, with the color scale denoting the corresponding IPR. At $W=W_{c1}=0.52$ [Fig.~\ref{fig:recom}(a)], the eigenvalues form several distinct branches with different localization characteristics. For $W=W_{c2}=1.42$ [Fig.~\ref{fig:recom}(b)], the spectrum is substantially rearranged but remains distributed over disconnected or open branches, while the intermediate case $W=1.30$ [Fig.~\ref{fig:recom}(c)] exhibits a qualitatively similar non-loop-like structure. Thus, the localization transition is accompanied by a redistribution of states within the complex-energy spectrum rather than by the emergence of a conventional closed spectral loop. In all three cases, eigenvalues occurring in imaginary plane near $\mathrm{Re}(E)=0$ exhibit relatively small IPR values, indicating an extended character and highlighting the strong variation of localization properties across different regions of the complex spectrum.

\textit{We do not assign a point-gap winding number to the spectra shown in Fig.~\ref{fig:recom}, because the eigenvalues at fixed model parameters do not form a closed spectral loop in the complex-energy plane}. A point-gap winding number is defined from a continuously parameterized closed spectral trajectory that encircles a reference energy $E_{\mathrm{B}}$. In our case, the eigenvalues are distributed over disconnected or open branches, with no closed spectral trajectory generated by the parameters used in the present calculation. Thus, constructing a loop by artificially connecting the real and imaginary eigenvalue components would not have a direct physical meaning and could produce a prescription-dependent result. Since the present calculation is instead aimed at characterizing the localization-delocalization transitions, we use the complex-energy spectrum to examine the distribution and localization character of the eigenstates rather than to define a winding invariant.


\bibliography{references}

\end{document}